\documentclass[conference]{IEEEtran}
\IEEEoverridecommandlockouts
\usepackage{cite}
\usepackage{amsmath,amssymb,amsfonts}
\usepackage{graphicx}
\usepackage{textcomp}
\usepackage{xcolor}
\usepackage{booktabs}
\usepackage{multirow}
\usepackage{url}
\usepackage{balance}
\usepackage{enumitem}
\usepackage{colortbl}
\usepackage{array}
\usepackage{pgfplots}
\pgfplotsset{compat=1.18}

\def\BibTeX{{\rm B\kern-.05em{\sc i\kern-.025em b}\kern-.08em
    T\kern-.1667em\lower.7ex\hbox{E}\kern-.125emX}}

\begin{document}

\title{Agent-Callable Feature Coverage:\\Measuring Software Readiness for AI Agents}

\author{\IEEEauthorblockN{Zedong Peng}
\IEEEauthorblockA{\textit{Department of Computer Science} \\
\textit{University of Montana}\\
Missoula, MT, USA \\
zedong.peng@umt.edu}
}

\maketitle

\begin{abstract}
AI agents already interact with graphical software through screenshot-based computer use; no GUI is truly off-limits anymore. The urgent question is therefore not \emph{whether} agents can operate software, but how well software supports them through structured, controllable channels. We formalize this need as the \textit{GUI--API parity principle}: every capability accessible to human users through a graphical interface should also be accessible to agents through a structured, callable interface, with appropriate safety metadata. To operationalize this principle, we introduce two contributions. First, \textit{Agent-Callable Feature Coverage} (ACFC), a product-level readiness metric that quantifies the extent to which a software system's human-facing capabilities are accessible to and usable by AI agents. Second, the \textit{Agent Readiness Conformance} (ARC), a multi-dimensional assessment framework that scores each capability on three axes: \textit{Accessibility} (whether an agent can call it), \textit{Discoverability} (whether an agent can find and understand it), and \textit{Controllability} (whether an agent can use it safely), each on a 0--3 scale, yielding a composite readiness score of 0--9 per capability. ARC draws on the Richardson Maturity Model and its multi-dimensional extensions to provide finer-grained assessment than binary coverage. Through an empirical study of 30 software systems across five categories, we find that ARC-based ACFC scores are substantially lower than Accessibility-only metrics would suggest, revealing that most sampled systems achieve moderate API coverage but lag on agent-oriented documentation and safety governance. In a controlled local testbed of ten deployed systems, API-only agents completed 56 of 58 tasks whose target capabilities were exposed through structured interfaces; a documentation ablation reduced this to 50 of 58, providing controlled evidence that Discoverability affects agent execution even among accessible capabilities. Inaccessible capabilities were included as negative controls and were not solved under API-only execution. The full-reference accessibility pattern is robust across three agent models, including Google's open-weights Gemma~4 run locally.
\end{abstract}


\section{Introduction}
\label{sec:intro}

The way people use software is changing. Users increasingly describe goals in natural language and delegate execution to AI agents: ``schedule this meeting,'' ``triage these support tickets,'' ``deploy the staging branch.'' What began as conversational assistance has evolved into autonomous task execution, with agents navigating applications, composing multi-step workflows, and completing operations that previously required manual interaction~\cite{sweagent2024, humantoagent2025, webarena2024, appworld2024}.

Computer-use capabilities have accelerated this shift dramatically. Agents that perceive screens through screenshots and manipulate interfaces through synthetic input can, in principle, operate \emph{any} graphical application~\cite{osworld2024, comact2025, seeaction2025, droidagent2024}. The GUI, long the de facto boundary between software and its users, no longer gates who, or what, can interact with it. Restricting software to GUI-only access does not prevent agents from using it; it merely forces them into fragile, uncontrolled screen-scraping.

This creates a new engineering challenge. Screen-based interaction is slow, error-prone, and operates outside the software's permission model, audit trail, and error-handling logic. An agent that clicks through a GUI holds no scoped credential, triggers no structured event, and cannot verify its actions beyond parsing the next screenshot. The result is a paradox: agents can technically reach any GUI feature, yet without structured interfaces this access is fragile and ungoverned. Meanwhile, even software that does provide APIs often exposes only a subset of its GUI capabilities. A video editing application may offer dozens of features through its interface but provide API access to just a handful; a project management tool may support rich dashboards visually yet offer no programmatic way to retrieve the underlying data. We term this gap between human-facing and agent-callable capabilities a \textit{feature parity gap}.

This situation parallels a challenge software engineering has faced before: accessibility for users with disabilities~\cite{reca11_2025, timestamp2025}. Before standards like WCAG and ARIA, most software was usable only by sighted, able-bodied users. The solution was not to build separate applications for disabled users, but to embed structured accessibility metadata (screen reader labels, semantic roles, keyboard navigation paths) into the same software. Similarly, we argue that software need not build entirely separate ``agent versions,'' but should ensure that its human-facing capabilities are also exposed through structured, callable interfaces.

Yet before we can improve agent accessibility, we must first \textit{measure} it, and establish what ``ready'' means beyond mere existence of an API endpoint.

We begin from a simple principle: \textbf{if a capability is accessible to a human through a GUI, it should also be accessible to an agent through a structured API}, equipped with appropriate safety metadata. We call this the \textit{GUI--API parity principle}. Now that computer-use agents have rendered GUIs permeable, this principle is no longer aspirational but urgent: agents \emph{will} interact with software regardless; the question is whether they do so through controlled, governed channels or through brittle screen simulation.

To operationalize this principle, we propose two complementary contributions. First, \textbf{Agent-Callable Feature Coverage (ACFC)}, a product-level readiness metric that measures how fully a software system's human-facing capabilities are exposed to agents through official, stable, structured interfaces such as REST APIs, MCP tools, SDK methods, CLI commands, or automation endpoints. ACFC is grounded in Jackson's requirements theory~\cite{jackson1997meaning}, which frames software capabilities as mediated by \textit{shared phenomena} between the machine and its environment.

Second, the \textbf{Agent Readiness Conformance (ARC)}, a multi-dimensional assessment framework inspired by the Richardson Maturity Model~\cite{richardson2008rmm} and its extensions~\cite{ws3model2015, maturitycube2022}. ARC evaluates each capability on three dimensions (Accessibility, Discoverability, and Controllability), each scored 0--3, yielding a composite readiness score of 0--9. This replaces the coarse binary (covered/uncovered) assessment with a graduated scale that captures not just whether an API exists, but whether an agent can find it, understand it, and use it safely.

We conduct an empirical study of 30 software systems across five categories (creative tools, productivity tools, developer tools, business tools, and open-source self-hosted applications) and address three research questions:

\begin{itemize}[leftmargin=*]
\item \textbf{RQ1 (Readiness landscape):} We measure ACFC distributions across software categories in our sample.
\item \textbf{RQ2 (Gap taxonomy):} We develop an empirically grounded taxonomy of feature parity gap types.
\item \textbf{RQ3 (Behavioral validation):} In a controlled testbed we test whether Accessibility is a substrate for execution and whether Discoverability affects success among accessible capabilities.
\end{itemize}

\noindent This paper makes the following contributions:
\begin{enumerate}[leftmargin=*]
\item We articulate the \textit{GUI--API parity principle} and define Callable Coverage (CC) and Agent-Callable Feature Coverage (ACFC) as complementary metrics for quantifying the gap between human-facing and agent-facing software interfaces.
\item To operationalize this principle, the \textit{Agent Readiness Conformance} (ARC) framework assesses per-capability agent readiness on three dimensions (Accessibility, Discoverability, Controllability), grounded in REST API maturity model literature.
\item Applying ARC to 30 software systems spanning five categories yields a dataset of capability-to-interface mappings and an empirical characterization of the prevalence and types of feature parity gaps.
\item A behavioral validation on ten locally deployed systems shows that API-only agents complete nearly all tasks whose capabilities have structured callable interfaces, and a controlled documentation ablation provides evidence that reduced Discoverability lowers success even when Accessibility is held fixed.
\end{enumerate}

\noindent All supplementary materials are available at \texttt{https://doi.org/10.5281/zenodo.21098126}.

\section{Background}
\label{sec:background}

\subsection{Requirements, Specifications, and Shared Phenomena}

Jackson's seminal work on the meaning of requirements~\cite{jackson1997meaning} establishes that software requirements describe desired states in the \textit{environment}, not behaviors of the machine itself. The machine influences the environment through \textit{shared phenomena}: events, states, or values observable by both the machine and external entities at the machine-environment boundary. A \textit{specification} prescribes how the machine should behave on these shared phenomena so that, combined with environmental assumptions, the requirements are satisfied: $E, S \vdash R$.

This framework directly informs ACFC. A software product's capabilities can be projected onto different shared phenomena surfaces for different user types. A GUI button click, a REST API call, and an MCP tool invocation are all shared phenomena through which a user (human or agent) interacts with the same underlying capability. ACFC measures whether the agent-facing projection covers the same capability set as the human-facing projection.

\subsection{Agent Interaction Modalities}

AI agents interact with software through several modalities, forming a spectrum from unstructured to fully structured:

\textbf{GUI automation} (browser agents~\cite{webarena2024}, mobile agents~\cite{droidagent2024}) simulates human interaction but is fragile and error-prone~\cite{seeaction2025, uxllm2025}.
\textbf{Structured APIs} (REST, GraphQL, SDK methods) provide stable, typed interfaces; API-based agents significantly outperform browser-only agents~\cite{beyondbrowsing2024}.
\textbf{Tool protocols} such as MCP~\cite{mcp2025} and OpenAPI~\cite{openapi2024} standardize how capabilities are exposed to AI systems.
\textbf{Hybrid approaches} combine GUI and API interaction per task~\cite{mcpworld2025, comact2025}.
The choice of modality matters~\cite{beyondbrowsing2024, comact2025, apiagentsgui2025}, motivating measurement of which capabilities are accessible through structured interfaces versus GUI-only.

\section{Conceptual Framework}
\label{sec:definitions}

\subsection{Human-Facing Capability}

\begin{quote}
\textbf{Definition 1.} A \textit{human-facing capability} is a goal-directed function that is advertised, documented, or made available to human users as a coherent unit of value through a software product's human-facing interface.
\end{quote}

Capabilities are defined at the \textit{user goal} level, not at the interface element level. ``Export video as MP4'' is a capability; ``click the export button'' is not. ``View analytics dashboard'' is a capability; ``scroll to chart widget'' is not. A capability must represent an independently valuable outcome that a user would recognize as a feature of the software.

\subsection{Agent-Callable Counterpart}

\begin{quote}
\textbf{Definition 2.} An \textit{agent-callable counterpart} of a human-facing capability is an official, stable, structured interface, such as an API operation, MCP tool, SDK method, CLI command, COM interface, or automation endpoint, that allows an AI agent to invoke an equivalent or sufficient form of that capability without relying on GUI simulation.
\end{quote}

Critically, GUI automation (e.g., Playwright, Selenium, computer-use agents) does \textit{not} qualify as an agent-callable counterpart. While agents can simulate human GUI interaction, such simulation does not represent the software intentionally exposing capabilities to programmatic users. Similarly, unofficial, undocumented, or unstable endpoints do not qualify.

\subsection{The Agent Readiness Conformance (ARC)}

Binary coverage (exists/absent) is too coarse for assessing agent readiness. An API endpoint may exist but be undocumented, undiscoverable, or lack safety controls. Inspired by how the Richardson Maturity Model (RMM)~\cite{richardson2008rmm} graduated REST API quality from Level~0 to Level~3, and how subsequent work extended RMM into multi-dimensional cubes combining REST maturity, data abstraction, and security~\cite{ws3model2015, maturitycube2022}, we propose the \textit{Agent Readiness Conformance} (ARC): three analytically separable but operationally ordered dimensions, each scored 0--3.

Each capability is scored 0--3 on three dimensions, \textbf{Accessibility} (A: whether an agent can call it), \textbf{Discoverability} (D: whether an agent can find and understand it), and \textbf{Controllability} (C: whether an agent can use it safely), following the rubric in Table~\ref{tab:arc_rubric}.

\begin{table*}[t]
\centering
\caption{The ARC rubric. Each human-facing capability is scored 0--3 independently on Accessibility (A), Discoverability (D), and Controllability (C); the composite is $\text{ARC}(f) = A(f)+D(f)+C(f) \in [0,9]$.}
\label{tab:arc_rubric}
\footnotesize
\begin{tabular}{@{}c p{0.30\textwidth} p{0.30\textwidth} p{0.30\textwidth}@{}}
\toprule
\textbf{Lvl} & \textbf{A: Accessibility} (callable) & \textbf{D: Discoverability} (findable, intelligible) & \textbf{C: Controllability} (safely usable) \\
\midrule
\textbf{0} & GUI-only; no API endpoint exists. & Undocumented, or human-only docs (PDF, wiki). & No safety metadata; basic authentication only. \\
\textbf{1} & API exists but incomplete (read-only, missing parameters, partial). & Machine-readable spec (OpenAPI/Swagger) with minimal descriptions. & Scoped authentication (OAuth scopes, permission-gated keys). \\
\textbf{2} & Full API coverage with correct HTTP semantics (CRUD, status codes). & Rich descriptions: schemas, examples, parameter constraints. & Risk annotations and operation classification (read/write/destructive). \\
\textbf{3} & API with pre/postconditions, idempotency, and verification endpoints. & Full affordance model: prompt hints, capability metadata, runtime discovery~\cite{amundsen2016wadm}. & Human-in-the-loop confirmation, recovery procedures, audit trails. \\
\bottomrule
\end{tabular}
\end{table*}

\smallskip
\noindent\textbf{Staged interpretation.} The three dimensions form a staged dependency chain: Accessibility provides the \textit{operational substrate} (without a callable interface, nothing can be documented or governed); Discoverability provides \textit{epistemic access} (the agent must identify and understand the operation); Controllability provides \textit{governed delegation} (safe autonomous invocation). Crucially, a permission boundary or risk label contributes to Controllability only insofar as it is exposed in a form the agent can parse and use for planning; Discoverability is therefore an epistemic prerequisite for effective Controllability. ARC should be read as a \textit{cube for diagnosis} (each dimension reveals a distinct gap type) but as a \textit{ladder for readiness} (call, then understand, then safely delegate).

The per-capability ARC score is: $\text{ARC}(f) = A(f) + D(f) + C(f) \in [0, 9]$. We use this additive form as a \emph{diagnostic} score that makes per-dimension bottlenecks visible, not as a claim that the dimensions are substitutable; Section~\ref{sec:discussion} evaluates a staged alternative that respects the A$\rightarrow$D$\rightarrow$C dependency.

\subsection{Coverage and Readiness Metrics}

We define two complementary metrics. The first captures binary accessibility:

\begin{quote}
\textbf{Definition 3a.} The \textit{Callable Coverage} of a software system $S$ is:
\begin{equation}
\text{CC}(S) = \frac{|\{f \in F_H(S) : A(f) \geq 2\}|}{|F_H(S)|}
\end{equation}
\end{quote}

\noindent CC measures the fraction of capabilities with at least full API coverage ($A \geq 2$), providing a binary accessibility check independent of documentation or safety quality.

The second integrates all three ARC dimensions into a readiness-weighted score:

\begin{quote}
\textbf{Definition 3b.} For a software system $S$ with human-facing capability set $F_H(S)$, the \textit{Agent-Callable Feature Coverage} (ACFC) is:
\begin{equation}
\text{ACFC}(S) = \frac{1}{|F_H(S)|} \sum_{f \in F_H(S)} \frac{\text{ARC}(f)}{9}
\end{equation}
\end{quote}

\noindent ACFC ranges from 0 to 1. Unlike CC, which only captures Accessibility, ACFC reflects the joint quality across all three ARC dimensions: a system with high CC but poor Discoverability or Controllability will have a lower ACFC than its CC alone would suggest. The complementary metric $1 - \text{ACFC}(S)$ gives the \textit{Feature Parity Gap Rate}.

In our empirical study (Section~\ref{sec:methodology}), we apply full three-dimensional ARC scoring: each capability is assessed on A, D, and C by an expert coder following a detailed coding guideline, and every score records the API or documentation evidence that justifies it.

\subsection{Feature Parity Gap}

\begin{quote}
\textbf{Definition 4.} A \textit{feature parity gap} occurs when a human-facing capability scores below maximum on any ARC dimension, lacking an adequate callable counterpart, having insufficient documentation, or missing safety controls.
\end{quote}

We identify five types of feature parity gaps, subject to empirical refinement (see Section~\ref{sec:methodology}):

\begin{itemize}[leftmargin=*]
\item \textbf{Missing projection (A0):} The capability has no API/tool counterpart at all.
\item \textbf{Partial projection (A1):} An API exists but covers only a subset of the capability's functionality.
\item \textbf{Granularity mismatch:} The API operates at a different abstraction level than the GUI.
\item \textbf{Semantic drift:} The API and GUI model the same capability differently.
\item \textbf{Verification gap:} The API can invoke the capability but provides insufficient evidence to confirm the outcome.
\end{itemize}

\section{Related Work}
\label{sec:related}

\subsection{Coverage Metrics and GUI-to-API Traceability}

Coverage metrics are a standard instrument for making otherwise invisible software gaps measurable, and prior work has extended them beyond source code. Restats~\cite{restats2021} supports eight REST API coverage criteria computed from HTTP traffic against OpenAPI specifications; Ticket Coverage~\cite{ticketcoverage2018} contextualizes test coverage around issue tickets; MBTCover~\cite{mbtcover2024} integrates requirements, model, and code coverage for industrial web testing; and AutoE2E~\cite{autoe2e2025} infers feature-level end-to-end tests, the closest work to measuring \emph{feature-level} coverage in web applications. In GUI testing, event-flow and event-dependency models measure interaction coverage~\cite{greyboxgui2013, guitracer2017}. A related line connects UI behavior to APIs: Carving UI Tests~\cite{carvingui2023} infers REST APIs from web UI navigation with 98\% precision. All these metrics measure whether \textit{tests} cover code, API operations, or GUI events. ACFC measures a different dimension: whether \textit{product-level human-facing capabilities} are covered by \textit{agent-callable interfaces}.

\subsection{Agent GUI, API, and Hybrid Interaction}

Agent research consistently shows that interaction modality affects task performance: API-based and hybrid agents outperform browser-only ones~\cite{beyondbrowsing2024, apiagentsgui2025}, and COM-based executable abstractions outperform GUI-only approaches for professional software~\cite{comact2025}. MCPWorld~\cite{mcpworld2025} explicitly studies how application features can be exposed as callable APIs for agents. In the SE domain, autonomous LLM agents have been deployed for program repair~\cite{repairagent2025, googlerepair2025, alignrepair2025}, code design issue localization~\cite{localizeagent2025}, and end-to-end industrial testing~\cite{autotest2025}, all relying on structured tool interfaces. Xia et al.~\cite{agentless2025} further argue that understanding when and why SE agents succeed or fail is prerequisite for trustworthy deployment. These studies evaluate agent \textit{trajectories and policies}; ACFC measures the \textit{software itself}: how much capability it structurally exposes to agents.

\subsection{API Maturity Models}

Richardson's REST Maturity Model (RMM)~\cite{richardson2008rmm} defined four levels that became the standard yardstick for REST API design quality. Multi-dimensional extensions add semantic, documentation, and security axes~\cite{ws3model2015, maturitycube2022, amundsen2016wadm}, though most production APIs cluster at Level~2~\cite{rodriguez2016restcompliance}. Hypermedia affordances enable autonomous agent navigation~\cite{vachtsevanou2023signifiers, verborgh2017pragmatic}, and LLMs can automatically derive RMM levels from OpenAPI specifications~\cite{smardas2025llmrmm}, suggesting partial automation of ARC scoring.

To our knowledge, no existing maturity model incorporates \textit{agent-specific} dimensions such as prompt-friendly descriptions, risk annotations, or human-in-the-loop governance. Industry frameworks have begun addressing this gap: Cloudflare's Agent Readiness framework~\cite{cloudflare2026agentreadiness} defines four dimensions, and Postman's API Assessment~\cite{postman2025stateofapi} includes an explicit ``AI Readiness'' dimension. However, to our knowledge, no peer-reviewed academic framework formalizes per-capability agent readiness across Accessibility, Discoverability, and Controllability. ARC fills this gap.

\subsection{Agent-Ready APIs and Tool Quality}

OpenAPI~\cite{openapi2024} and MCP~\cite{mcp2025} provide infrastructure for machine-readable interfaces, but structural availability does not guarantee agent usability. Structurally valid documentation still causes agent planning failures~\cite{openapismells2025}; MCP tool description quality affects agent behavior in non-obvious ways~\cite{mcpsmells2025}; and automated OpenAPI-to-MCP conversion surfaces specification inconsistencies~\cite{automcp2025}. Multi-agent and LLM-driven REST testing~\cite{multiagentrest2025, autoresttest2025, llamaresttest2025} further confirms that documentation completeness directly affects test generation quality. Collectively, these findings motivate ARC's three dimensions: Accessibility alone is insufficient without Discoverability and Controllability.

\section{Methodology}
\label{sec:methodology}

\subsection{Software Sample Selection}

We select 30 software systems across five categories to ensure diversity in domain, interface maturity, and expected ACFC levels:

\begin{itemize}[leftmargin=*]
\item \textbf{Creative tools} (6): Software for design, media editing, and visual creation, expected to have rich GUI capabilities with limited API coverage.
\item \textbf{Productivity tools} (6): Document, task, and knowledge management tools, expected to have moderate API coverage.
\item \textbf{Developer tools} (6): Code hosting, issue tracking, and CI/CD platforms, expected to have high API coverage (serving as a high-ACFC reference group).
\item \textbf{Business tools} (6): E-commerce, CRM, and payment platforms, expected to have extensive APIs with potential governance and workflow gaps.
\item \textbf{Open-source self-hosted} (6): Self-deployable applications with public source code, API documentation, and feature lists.
\end{itemize}

Selection criteria require each software to have: (1) publicly accessible feature documentation or product pages, (2) publicly accessible API or developer documentation, and (3) sufficient maturity and user base to provide a mature, widely used exemplar within its category. The 30 systems, their primary agent-callable interfaces, and their ARC-based scores are listed together in Table~\ref{tab:acfc_scores}.

\subsection{Capability Extraction Pipeline}

We employ a semi-automated pipeline combining LLM-assisted extraction with human validation:

\textbf{Step 1: Automated candidate extraction.} For each software system, we collect official feature pages, help documentation, product tours, and changelogs. We prompt a large language model to extract candidate capabilities, constrained to functions that are (a) goal-directed, (b) user-visible, and (c) independently valuable. The prompt instructs the model to distinguish capabilities from UI operations and technical endpoints.

\textbf{Step 2: Human validation.} An expert reviewer checks each candidate list against the coding guideline, merging overlapping candidates, splitting compound capabilities, and removing entries that do not meet the definition. We target 30--50 validated capabilities per software system.

\textbf{Step 3: Granularity calibration.} A capability must be expressible as an independent user goal (e.g., ``export project data'') rather than a UI action (``click export button'') or an internal operation (``write to database''). If a complex workflow contains multiple independently valuable sub-goals, it is decomposed accordingly.

Every automatically extracted candidate is manually reviewed against the guideline before scoring, and we release the complete validated capability lists for inspection.

\subsection{ARC Scoring Protocol}

For each validated capability, we search the software's official API documentation, MCP tool registry, SDK reference, and CLI documentation, and score the capability on all three ARC dimensions following a detailed coding guideline (available in the replication package).

We score each dimension on the 0--3 rubric of Table~\ref{tab:arc_rubric}, judging Accessibility from whether and how completely an agent-callable interface covers the GUI capability, Discoverability from the machine-readability and semantic richness of its documentation, and Controllability from the operation-level safety and governance metadata exposed to the caller. When A=0, the remaining dimensions are scored 0 by definition.

\textbf{Exclusions.} GUI automation tools (Selenium, Playwright, computer-use APIs) do not count as counterparts. Unofficial, undocumented, or deprecated endpoints are excluded. Third-party wrappers (e.g., community-built MCP servers) are excluded unless officially endorsed by the vendor.

\textbf{Gap classification.} For each capability scoring below ARC 9 on any dimension, we classify the gap type. Beyond the five Accessibility-focused types in Section~\ref{sec:definitions}, we additionally classify Discoverability gaps (D0--D2) and Controllability gaps (C0--C2) to characterize the full readiness profile.

\subsection{Coding Reliability}

Rather than an inter-rater statistic, our reliability argument rests on transparency and traceability. Each capability is scored by an expert coder against a detailed written guideline (released in the replication package), and \emph{every} A, D, and C score records the specific API endpoint or documentation passage that justifies it. We release the full per-capability scores together with this evidence for all 1{,}299 capabilities, so that any reader can inspect, challenge, or re-score the mapping.

\subsection{Granularity}

Because ACFC is a mean over capabilities, it depends on extraction granularity. We control this by defining capabilities at the user-goal level (Step~3 above) and targeting a comparable 30--50 capabilities per system across all software; a quantitative sensitivity study across coarser and finer granularities is left to future work.

\subsection{Agent Task Validation (RQ3)}
\label{sec:rq3method}

To assess whether ARC's dimensions correspond to observable constraints on API-based agent execution, we design a controlled experiment comparing agent task performance across capability coverage levels.

\textbf{Testbed.} We deploy ten open-source software systems locally, spanning the measured ACFC range. Three systems (Gitea, BookStack, Nextcloud) are drawn from the 30-system survey sample; the remaining seven (Snipe-IT, Grocy, Paperless-NGX, Immich, Vikunja, HedgeDoc, Firefly~III) were selected for local deployability and ACFC range diversity, with their ARC scores computed using the same pipeline (Section~\ref{sec:validation}). For each system we author 5--12 concrete tasks (78 total) spanning the ARC score range. Tasks target capabilities at different ARC levels, from A0 (no API) through A2--A3 (full API with varying D and C scores).

\textbf{State-based oracle.} Following Jackson's shared-phenomena view, we define success not by the agent's self-report but by the \emph{required environment state}: a task succeeds iff the target state holds when verified at the environment side (API read-back plus a direct database query). Crucially, each oracle keys only on attributes named in the task goal (entity names, stated values), never on incidental identifiers, so it accepts any correct solution and is robust to how the agent achieves it. For A=0 tasks, the database query verifies whether the target state was nevertheless reached through any available API path, preventing us from counting agent self-reports as failures or successes without checking the environment state.

\textbf{Agent configuration.} The agent operates black-box: it is given the task goal in natural language and the system's API reference, and acts solely through a thin HTTP tool that logs every call. It never sees the oracle, the reference solution, or the database. If no documented endpoint can accomplish a task, it is instructed to report inability. We run three agent models spanning the capability spectrum: Opus~4.8 (primary), Sonnet~4.6, and Google's Gemma~4 (31B, an open-weights model served locally via Ollama), to confirm the effects are not artifacts of a single model.

\textbf{Primary metric:} binary task success (completed/failed), determined solely by the oracle.

\textbf{Hypotheses:}
\begin{itemize}[leftmargin=*]
\item $H_{3a}$ (Accessibility as substrate): API-only agents can complete tasks whose target capabilities are exposed through structured interfaces ($A \geq 2$). Tasks with no official callable counterpart ($A=0$) are included as \emph{negative controls} for the task classification and oracle, not as a stochastic performance comparison.
\item $H_{3b}$ (Discoverability): among $A \geq 2$ tasks, reducing Discoverability lowers task success, tested by a controlled documentation ablation that holds Accessibility and the task fixed.
\end{itemize}

\textbf{Statistical analysis.} Our one inferential test is the Discoverability ablation ($H_{3b}$): we compare covered-task success under the full versus minimal reference with McNemar's exact test on the discordant pairs. For the $A=0$ negative controls we report counts descriptively; because the agent is restricted to official APIs and is instructed to report inability when no documented endpoint applies, we do not treat $A=0$ versus $A\geq2$ as a stochastic comparison. As a secondary, deliberately underpowered check we report the Spearman correlation between per-system Callable Coverage and success rate ($n=10$).

\section{Results}
\label{sec:results}

\subsection{RQ1: Agent Readiness Landscape}

Table~\ref{tab:acfc_scores} lists all 30 systems, their primary agent-callable interfaces, and their ARC-based ACFC scores. Each capability is scored on three dimensions (A, D, C each 0--3); ACFC is the mean of per-capability normalized ARC scores (ARC/9). The table also reports median scores on each dimension, revealing where readiness gaps concentrate.

\begin{table*}[t]
\centering
\caption{The 30-system sample and their ARC-based ACFC scores. ``Primary Interface Types'' are the official agent-callable surfaces considered. $\bar{A}$, $\bar{D}$, $\bar{C}$: per-capability median on each ARC dimension; $|F|$: total capabilities scored. Rows are ordered by ACFC within each category.}
\label{tab:acfc_scores}
\footnotesize
\begin{tabular}{@{}ll p{4cm} r rrr r@{}}
\toprule
\textbf{Category} & \textbf{Software} & \textbf{Primary Interface Types} & \textbf{ACFC} & $\bar{A}$ & $\bar{D}$ & $\bar{C}$ & $|F|$ \\
\midrule
\multirow{6}{*}{Creative}
 & Canva & Connect API (REST), MCP server & 0.52 & 2 & 2 & 2 & 51 \\
 & Miro & REST API v2 & 0.47 & 2 & 2 & 2 & 43 \\
 & Figma & REST API, Plugin API & 0.43 & 2 & 2 & 1 & 45 \\
 & Blender & Python API (bpy), CLI & 0.40 & 2 & 2 & 0 & 44 \\
 & DaVinci Resolve & Scripting API (Python/Lua) & 0.24 & 1 & 1 & 1 & 42 \\
 & CapCut & Open Platform (plugins only) & 0.04 & 0 & 0 & 0 & 45 \\
\cmidrule{2-8}
 & \textit{Category median} & & 0.41 & 2 & 2 & 1 & \\
\midrule
\multirow{6}{*}{Productivity}
 & Slack & Web API, Events API, CLI & 0.61 & 2 & 2 & 1 & 44 \\
 & Asana & REST API & 0.55 & 2 & 2 & 1 & 41 \\
 & Todoist & REST API v2, Sync API & 0.55 & 2 & 2 & 1 & 46 \\
 & Trello & REST API & 0.48 & 2 & 2 & 1 & 42 \\
 & Airtable & REST API, Metadata API & 0.47 & 1 & 2 & 2 & 42 \\
 & Notion & REST API, MCP server & 0.35 & 1 & 1 & 1 & 45 \\
\cmidrule{2-8}
 & \textit{Category median} & & 0.51 & 2 & 2 & 1 & \\
\midrule
\multirow{6}{*}{Developer}
 & GitHub & REST, GraphQL, gh CLI & 0.79 & 3 & 2 & 3 & 47 \\
 & GitLab & REST, GraphQL, glab CLI & 0.71 & 3 & 2 & 2 & 45 \\
 & Jira & REST API v3 & 0.69 & 2 & 2 & 2 & 41 \\
 & Vercel & REST API, vercel CLI & 0.69 & 2 & 2 & 2 & 45 \\
 & Linear & GraphQL API & 0.67 & 2 & 2 & 1.5 & 40 \\
 & Sentry & REST API, sentry-cli & 0.65 & 2 & 2 & 2 & 43 \\
\cmidrule{2-8}
 & \textit{Category median} & & 0.69 & 2 & 2 & 2 & \\
\midrule
\multirow{6}{*}{Business}
 & Stripe & REST API, Stripe CLI & 0.77 & 3 & 2 & 2 & 42 \\
 & HubSpot & REST API v3 & 0.69 & 2 & 2 & 2 & 42 \\
 & Zendesk & REST API, Zendesk CLI & 0.67 & 2 & 2 & 2 & 42 \\
 & Shopify & GraphQL Admin API, REST & 0.66 & 2 & 2 & 2 & 43 \\
 & Mailchimp & Marketing API v3 & 0.56 & 2 & 2 & 2 & 40 \\
 & QuickBooks Online & REST API (Intuit) & 0.54 & 2 & 2 & 2 & 44 \\
\cmidrule{2-8}
 & \textit{Category median} & & 0.67 & 2 & 2 & 2 & \\
\midrule
\multirow{6}{*}{Open-Source}
 & Outline & REST API & 0.59 & 2 & 2 & 1 & 43 \\
 & Gitea & Swagger REST API, tea CLI & 0.58 & 2 & 2 & 2 & 46 \\
 & Nextcloud & OCS REST, WebDAV/CalDAV & 0.58 & 2 & 1 & 2 & 43 \\
 & ERPNext & Frappe REST API & 0.53 & 2 & 2 & 1 & 42 \\
 & Plane & REST API & 0.42 & 2 & 1 & 1 & 39 \\
 & BookStack & REST API & 0.38 & 1 & 1 & 1 & 42 \\
\cmidrule{2-8}
 & \textit{Category median} & & 0.55 & 2 & 1.5 & 1 & \\
\midrule
\textbf{Overall} & & & \textbf{0.55} & 2 & 2 & 2 & \\
\bottomrule
\end{tabular}
\end{table*}

\textbf{Overall readiness.} The overall median ACFC across all 30 systems is 0.55, with scores ranging from 0.04 (CapCut) to 0.79 (GitHub). By contrast, binary Callable Coverage (CC) across all 1,299 capabilities is 71.5\% (929 capabilities score $A \geq 2$). The gap between CC and ACFC quantifies the readiness deficit from Discoverability and Controllability shortfalls: most accessible APIs exist but lack agent-oriented documentation or safety governance. Only 49 of 1,299 capabilities (3.8\%) achieve the maximum ARC score of 9, while 156 (12.0\%) remain at ARC~=~0.

\textbf{Category differences.} Developer tools lead with a median ACFC of 0.69, followed by Business (0.67), Open-Source (0.55), Productivity (0.51), and Creative (0.41). The gap between Developer/Business and Creative categories reflects fundamentally different design philosophies: developer tools were built API-first to serve integration ecosystems, while creative tools prioritize visual, spatial interactions that resist API decomposition.

\begin{figure}[t]
\centering
\begin{tikzpicture}
\begin{axis}[
    ybar,
    bar width=4pt,
    width=\columnwidth,
    height=4.5cm,
    ylabel={Mean ARC Score (0--3)},
    ymin=0, ymax=3.2,
    ytick={0,1,2,3},
    symbolic x coords={Creative,Productivity,Open-Source,Business,Developer},
    xtick=data,
    x tick label style={rotate=25, anchor=east, font=\scriptsize},
    legend style={at={(0.5,1.02)},anchor=south,legend columns=3,font=\scriptsize},
    ymajorgrids=true,
    grid style={dashed,gray!30},
    enlarge x limits=0.15,
]
\addplot[fill=blue!60] coordinates {(Creative,1.21) (Productivity,1.66) (Open-Source,1.84) (Business,2.11) (Developer,2.26)};
\addplot[fill=orange!60] coordinates {(Creative,1.14) (Productivity,1.60) (Open-Source,1.51) (Business,1.92) (Developer,2.05)};
\addplot[fill=red!50] coordinates {(Creative,0.81) (Productivity,1.25) (Open-Source,1.30) (Business,1.79) (Developer,2.00)};
\legend{A (Accessibility), D (Discoverability), C (Controllability)}
\end{axis}
\end{tikzpicture}
\caption{Mean ARC dimension scores by category. Controllability (C) consistently trails Accessibility (A) and Discoverability (D), confirming the ``ADC imbalance.'' Only Developer and Business categories maintain balanced profiles.}
\label{fig:arc_dims}
\end{figure}
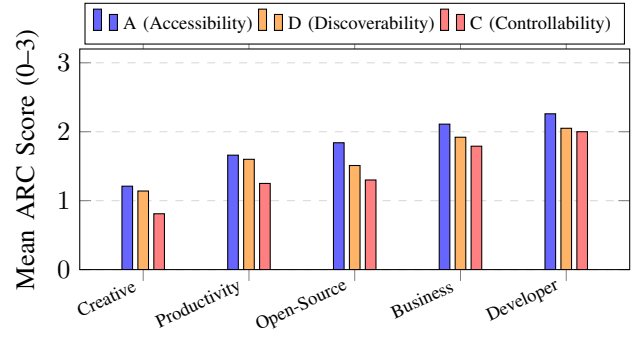

\textbf{Dimension-level analysis.} ARC's three-dimensional scoring reveals that Controllability is the primary readiness bottleneck. Most systems achieve moderate Accessibility ($\bar{A} = 2$): 71.5\% of capabilities score $A \geq 2$. Documentation is generally available at the D2 level ($\bar{D} = 2$). However, safety governance lags: only Developer and Business categories achieve a category-level $\bar{C} = 2$; Creative, Productivity, and Open-Source systems cluster at $\bar{C} = 1$. This ``ADC imbalance'' confirms that mere API existence understates the true readiness gap. Figure~\ref{fig:arc_dims} visualizes this pattern.

\textbf{Design philosophy and ACFC.} The correlation between software design philosophy and ACFC is striking. API-first platforms (GitHub, Stripe, GitLab) consistently achieve the highest ACFC scores (0.69--0.79), while GUI-first creative tools (CapCut, DaVinci Resolve) cluster at the bottom (0.04--0.24). However, the relationship is not strictly binary: Canva (0.52) and Miro (0.47) demonstrate that creative tools can achieve moderate ACFC by investing in REST APIs, even when their core value proposition is visual. Similarly, Notion (0.35) underperforms relative to other productivity tools despite significant developer community investment, because its API is narrowly focused on database operations while GUI-exclusive features (AI assistant, comments, permissions management) remain inaccessible. These patterns suggest that ACFC is driven not merely by whether APIs exist, but by how comprehensively they mirror the GUI's capability surface.

\textbf{Outliers and design philosophy.} GitHub (0.79) achieves the highest ACFC, driven by $\bar{A} = 3$ and $\bar{C} = 3$: its REST and GraphQL APIs include idempotency, status verification, fine-grained OAuth scopes, and approval workflows. Stripe (0.77) demonstrates that API-first design extends to safety governance: test mode, idempotency keys, and explicit risk annotations lift it above peers. At the other extreme, CapCut (0.04) exposes virtually no capabilities through structured APIs; its plugin platform offers no equivalent to its core video editing functions. Blender ($\bar{C} = 0$) illustrates that high Accessibility alone does not guarantee readiness: its Python API covers nearly all features but provides no scoped authentication or risk annotations, underscoring the value of ARC's multi-dimensional assessment.

\subsection{RQ2: Gap Taxonomy}

ARC's three-dimensional scoring reveals readiness gaps that binary coverage metrics cannot distinguish. We classify gaps on each dimension separately, then analyze the joint distribution.

\begin{table}[h]
\centering
\caption{Readiness Gap Distribution by ARC Dimension}
\label{tab:gap_types}
\footnotesize
\begin{tabular}{ll rr}
\toprule
\textbf{Dim.} & \textbf{Gap Type} & \textbf{Count} & \textbf{\%} \\
\midrule
\multirow{2}{*}{A} & A0: Missing projection     & 156 & 12.0\% \\
                    & A1: Partial projection     & 214 & 16.5\% \\
\midrule
\multirow{2}{*}{D$^*$} & D0--D1: Undiscoverable/minimal & 275 & 24.1\% \\
                    & D2: Rich but not agent-oriented & 721 & 63.1\% \\
\midrule
\multirow{3}{*}{C$^*$} & C0: No safety metadata     & 83 & 7.3\% \\
                    & C1: Scoped auth only       & 407 & 35.6\% \\
                    & C2: Risk-annotated, no HITL & 519 & 45.4\% \\
\bottomrule
\multicolumn{4}{l}{\scriptsize $^*$Among 1,143 capabilities with $A \geq 1$ ($n{=}1{,}299$ total).} \\
\end{tabular}
\end{table}

\textbf{Accessibility gaps.} Missing (A0) and Partial (A1) projections affect 370 capabilities (28.5\% of 1,299). A0 gaps (156, 12.0\%) concentrate in creative tools (CapCut, DaVinci Resolve) and GUI-heavy features of productivity tools (e.g., Notion's AI features, Airtable's formula builder). A1 gaps (214, 16.5\%) primarily involve read-only APIs that lack write capabilities, missing parameters, or subset functionality.

\textbf{Discoverability gaps.} Among 1,143 capabilities with API access ($A \geq 1$), 275 (24.1\%) score D0--D1: machine-readable specs either do not exist or provide only minimal descriptions. More strikingly, 721 capabilities (63.1\% of accessible) sit at the ``D2 ceiling'', i.e., rich human-oriented documentation with schemas and examples, but lacking agent-oriented affordances such as prompt hints, semantic operation descriptions, or runtime capability discovery. Only 147 capabilities (12.9\%) achieve D3.

\textbf{Controllability gaps.} Controllability is the most pervasive bottleneck. Among accessible capabilities, 490 (42.9\%) score C0--C1: either no operation-level safety controls beyond basic authentication (C0: 83) or scoped authentication without risk annotations (C1: 407). The ``C1 ceiling'' (APIs with permission scopes but no operation classification, audit events, or human-in-the-loop governance) is the single most common gap pattern. Only 134 capabilities (11.7\%) achieve C3 with full governance.

\textbf{The readiness bottleneck.} The joint analysis confirms that Controllability is the primary bottleneck: $\bar{C}_{mean} = 1.42$ trails both $\bar{A}_{mean} = 1.81$ and $\bar{D}_{mean} = 1.64$. The ``CRUD ceiling'' observed in binary coverage studies manifests as an ``ADC imbalance'': APIs exist and are documented, but lack the safety metadata that responsible agent operation requires.

\textbf{Cross-dimension gap patterns.} Three profiles dominate. \textit{A2/D2/C1 (``Accessible but ungoverned'')} is the most common (${\sim}$35\% of A$\geq$2 capabilities): full API and documentation but no operation-level safety beyond OAuth. \textit{A2/D1/C0 (``Available but opaque'')}: working APIs with minimal documentation and no safety metadata (common in Blender, ERPNext). \textit{A1/D2/C1 (``Partially covered'')}: API covers only a subset of GUI functionality despite good documentation (Notion, Airtable, BookStack). These profiles suggest that improving readiness requires coordinated investment across dimensions.

\section{Validation}
\label{sec:validation}

\subsection{RQ3: Behavioral Validation of ARC Dimensions}

To assess whether ARC's dimensions correspond to observable constraints on API-based agent execution, we conducted a controlled, fully reproducible agent task experiment on ten locally deployed open-source systems spanning a wide range of callable-coverage levels: Gitea, BookStack, Nextcloud, Snipe-IT, Grocy, Paperless-NGX, Immich, Vikunja, HedgeDoc, and Firefly~III. We do not evaluate ACFC as a scalar performance predictor; RQ3 instead tests two behavioral implications of ARC: that structured Accessibility makes API-only execution feasible, and that reduced Discoverability lowers success among already-accessible capabilities. Three systems (Gitea, BookStack, Nextcloud) carry full ARC scores from the 30-system survey; for the other seven we report Callable Coverage computed from the same capability inventories.

\textbf{Setup.} For each system we author 5--12 tasks spanning ARC score levels, with success defined by a state-based oracle verified at the environment side (Section~\ref{sec:rq3method}). The black-box agent receives only each task goal and the system's API reference, acting through a logged HTTP tool. We run three agent models: Opus~4.8 (primary), Sonnet~4.6, and Google's Gemma~4 (31B, open-weights, via Ollama).\footnote{Gemma~4 configuration: Ollama tag \texttt{gemma4:31b}; decoding temperature 0, context window 8192; driven by a ReAct-style text protocol (one \texttt{CALL <method> <path> [body]} action per turn) with a 15-turn budget per task. The full driver and per-task transcripts are in the replication package.}

\textbf{$H_{3a}$: Accessibility is a usable substrate.} Under the full API reference, the agent completed 56 of 58 tasks whose target capabilities were exposed through structured interfaces ($A \geq 2$), with success judged by the state-based oracle rather than agent self-report. Covered-task completion was high across all ten systems, including the lowest-coverage ones (HedgeDoc, BookStack), so structured callable interfaces make task execution broadly feasible for API-only agents (the per-system breakdown is in the replication package). The two failures were not caused by lack of Accessibility (analyzed below). Table~\ref{tab:rq3conditions} summarizes the three conditions. One BookStack task (revert a page to a prior revision) was excluded post hoc as an invalid $A{=}0$ instance: its required end state (page content equal to the original) is reachable through the covered page-edit API, so it does not isolate a no-API capability.

\begin{table}[h]
\centering
\caption{RQ3 conditions (Opus~4.8). Accessibility is treated as a substrate and the $A=0$ tasks as negative controls; the Discoverability ablation is the one inferential test.}
\label{tab:rq3conditions}
\footnotesize
\begin{tabular}{@{}l r r l@{}}
\toprule
\textbf{Condition} & \textbf{Tasks} & \textbf{Success} & \textbf{Role} \\
\midrule
$A\geq2$, full reference    & 58 & 56/58 & Main execution result \\
$A\geq2$, minimal reference & 58 & 50/58 & Discoverability ablation \\
$A=0$, API-only            & 19 & 0/19  & Negative control \\
\bottomrule
\end{tabular}
\end{table}

\textbf{$H_{3b}$: Discoverability affects execution (controlled ablation).} Because every system receives an equally curated reference, that design holds documentation quality constant and cannot isolate Discoverability. We therefore re-run each covered task under a \emph{minimal} reference (endpoint method and path only, stripped of parameter names, schemas, examples, and prose), holding Accessibility and the task fixed. Covered-task success falls from 56/58 to 50/58; the effect is monotone and significant (six tasks regress, none improve; McNemar's exact test $p = 0.031$). The regressions concentrate in capabilities with non-obvious request payloads (four Snipe-IT asset/checkout operations whose required identifier fields are not guessable, the Immich asset upload, and a Firefly~III rule), whereas operations with conventional REST bodies (creating named entities in Grocy, Gitea, Nextcloud, Vikunja, Paperless) are unaffected. Unlike the accessibility contrast, this result is not structurally predetermined: it is direct, controlled evidence that Discoverability shapes agent success once a capability is callable. This ablation is intentionally severe: it probes a lower-bound form of Discoverability rather than modeling all real-world D1 documentation, establishing that documentation content has a causal effect under fixed Accessibility; finer-grained D1/D2/D3 ablations are left to future work.

\textbf{Negative controls ($A=0$).} We also ran 19 tasks whose target capabilities have no official callable counterpart; none was completed under the API-only setting. This is expected by construction: with no documented endpoint, and with the agent instructed to report inability rather than improvise, such tasks are unsolvable through official APIs. We therefore use them as a sanity check on the task classification and the state oracle, confirming that the agent neither fabricated success nor reached the target state by an unintended path, rather than as evidence that ACFC predicts performance.\footnote{For completeness, the $A\geq2$ versus $A=0$ contingency is $[[56,2],[0,19]]$ (Fisher's exact $p<10^{-15}$); we report it descriptively and do not rely on it inferentially.} As a secondary, deliberately underpowered check, the per-system Spearman correlation between Callable Coverage and success rate is positive but not significant ($\rho = 0.38$, $n = 10$), attenuated because $A=0$ tasks are over-sampled and two systems contribute only accessible tasks.

\textbf{Failure analysis.} The two covered failures both occur in well-documented ($\bar{D}{=}2$) systems and are unrelated to documentation. \emph{PL-C4} (Paperless saved view): the agent created the view but did not populate all required sub-attributes, a task-decomposition error. \emph{IM-C5} (Immich system configuration): the API requires a full-object \texttt{PUT}, and the agent could not assemble a valid complete payload, drawing four successive HTTP~400 responses, an API-ergonomics failure. Accessibility ($A\geq2$) is thus a necessary substrate but not a sufficient condition; residual failures stem from task complexity and payload ergonomics.

\textbf{Cross-model robustness.} We repeat the full-reference experiment with the other two agents (Sonnet~4.6 and Gemma~4). The accessibility gate is invariant across all three: every one of the 19 negative controls is unsolved by every model (0/19 each). Covered-task completion, by contrast, scales with agent capability: 56/58 (Opus~4.8), 55/58 (Sonnet~4.6), and 45/58 (Gemma~4). That even a locally-run open-weights model completes most tasks whose capabilities are exposed through structured interfaces reinforces that Accessibility is a necessary substrate, while its lower completion rate confirms that using that substrate effectively still depends on the agent. The two hardest covered tasks (PL-C4, IM-C5) fail for all three models.

\section{Discussion}
\label{sec:discussion}

\subsection{Implications for Software Engineering}

Software in our sample exhibits a systematic ``CRUD ceiling'': operational capabilities are well-exposed through APIs, while analytical, configurational, and integrative capabilities remain GUI-exclusive. This pattern reflects historical integration needs, not deliberate design. The emergence of AI agents changes this calculus: an agent tasked with ``set up this project management tool, connect it to GitHub, and report weekly velocity'' encounters feature parity gaps at every step beyond basic issue creation. The ARC framework provides a structured path for closing these gaps dimension by dimension.

The CRUD ceiling manifests differently across categories. In creative tools, the gap is fundamental: spatial design operations resist decomposition into discrete API calls (Blender's Python API achieves high Accessibility but $\bar{C} = 0$, illustrating the challenge of retrofitting safety governance). In productivity tools, vendors expose core CRUD operations but gate analytical and administrative features behind the GUI. In business tools, individual operations are well-exposed, but multi-step approval flows often remain GUI-configured.

These patterns suggest a prioritized remediation path. For creative tools, developing richer API abstractions that capture spatial and temporal operations is the primary challenge. For productivity and business tools, extending existing APIs to cover administrative and analytical features is more tractable. Across all categories, adding Controllability metadata (risk annotations, operation classification, human-in-the-loop governance) to already-accessible operations represents a natural first step, as it addresses the most pervasive bottleneck without requiring new API development.

\subsection{Toward AI-Accessible Software}

Just as the web accessibility movement led to WCAG standards and accessibility audits, we envision ACFC and ARC as foundations for \textit{AI accessibility} assessment. The GUI--API parity principle implies that software vendors should treat agent accessibility as a first-class design concern, not an afterthought. Computer-use agents already bypass GUI restrictions through screen-scraping~\cite{osworld2024}; rather than accepting this fragile workaround, software should proactively expose controlled API channels with safety metadata.

The parallel with web accessibility extends to specific mechanisms. WCAG~2.0's four principles (Perceivable, Operable, Understandable, Robust) map onto ARC's dimensions: Accessibility corresponds to Perceivable and Operable (detecting and invoking the capability), Discoverability to Understandable (comprehending what the capability does and how to use it), and Controllability to Robust (using it safely across different agent models and contexts). Just as WCAG introduced conformance levels (A, AA, AAA), ARC's 0--3 scales provide graduated targets that software vendors can adopt incrementally.

The ``curb cut effect'' also applies: APIs designed for agent accessibility benefit all programmatic consumers, including CI/CD pipelines, testing frameworks, and third-party integrations. Investments in Discoverability improve developer experience alongside agent usability, and Controllability investments (risk annotations, audit trails, human-in-the-loop governance) serve compliance and regulatory needs that extend well beyond AI agents, aligning the business case for agent readiness with broader enterprise requirements.

\subsection{Two Failure Modes of GUI--API Parity}

The GUI--API parity principle has a dark mirror. The failure ACFC primarily measures is \emph{under-exposure}: capabilities humans can reach but agents cannot. Our experiment also surfaces the opposite failure, \emph{under-governance}: capabilities that agents \emph{can} reach but that were never designed to be invoked autonomously. Two observations motivate this. First, goal states that designers gate behind GUI-only flows are often reachable by composing authorized API operations: the BookStack task we excluded from RQ3 (revert a page to a prior revision, a GUI-only feature with no revert endpoint) was nonetheless achieved by the agent through ordinary page-edit calls, reaching the intended end state without the intended capability. Second, agents enumerate whatever is machine-readable: in failed attempts on undocumented features, agents fetched the instance's OpenAPI document (\texttt{/docs.json}) to discover endpoints directly. The agent-reachable surface is thus bounded not by what a vendor intends to expose, but by what is callable and discoverable. This matters because the most common profile among accessible capabilities is A2/D2/C1 (``accessible and documented but ungoverned''; Section~\ref{sec:results}): operations reachable by agents yet carrying no operation-level risk classification, confirmation, or audit metadata. Treating the GUI as an implicit safety boundary is therefore unsafe once a callable, discoverable API exists beneath it, which reframes ARC's Controllability dimension from a readiness convenience into a safety requirement. Because our agents held authorized administrator credentials, we report this as a \emph{structural} risk rather than a measured vulnerability rate; systematically characterizing reachable-but-ungoverned operations is important future work. Our Controllability scoring reflects each software system's own exposed safety infrastructure rather than an external security evaluation; a dedicated security analysis would complement this work but lies outside our scope.

\subsection{Toward Automated ARC Assessment}

Manual ARC scoring is labor-intensive (~3,900 individual scores for 30 systems). Since LLMs can already derive Richardson maturity levels from OpenAPI specifications~\cite{smardas2025llmrmm}, a semi-automated ARC pipeline is feasible: checking endpoint existence for Accessibility, analyzing description richness for Discoverability~\cite{llamaresttest2025, autoresttest2025}, and inspecting authentication scopes for Controllability.

\subsection{Implications for Agent Framework Design}

Our findings carry implications for agent framework designers~\cite{agentless2025, agentictrust2025}. Current frameworks assume tools are well-documented and complete, but 28.5\% of capabilities lack adequate API counterparts and 24.1\% of accessible capabilities have minimal documentation. Frameworks should incorporate graceful degradation for partial API coverage (A1), routing to human fallback~\cite{hula2025} when full equivalence is unavailable; documentation quality assessment~\cite{codegenerrors2025} before tool use; and risk-aware planning that defaults to conservative behavior on C0--C1 endpoints.

\subsection{Staged vs.\ Additive Scoring}

The current ACFC formula treats A, D, and C as additive: $\text{ARC}(f) = A + D + C$, making per-dimension bottlenecks visible. The staged dependency chain (A$\rightarrow$D$\rightarrow$C) suggests an alternative: $\text{ARC}_{\text{stage}}(f) = A_f + \min(A_f, D_f) + \min(A_f, D_f, C_f)$, which penalizes ``hollow'' profiles where later-stage scores exceed earlier-stage prerequisites (e.g., $A\!=\!3, D\!=\!0, C\!=\!3$ yields 3 rather than 6). We retain the additive formula because our empirical data already exhibits the A$\geq$D$\geq$C pattern ($\bar{A} = 1.81$, $\bar{D} = 1.64$, $\bar{C} = 1.42$), making the practical difference small. The staged formulation merits investigation in future work.

\subsection{Threats to Validity}

\textbf{Construct validity.} Capability definitions involve human judgment. We mitigate this through an explicit written coding guideline, evidence recorded for every A, D, and C score, and public release of the full per-capability mapping for independent inspection.

\textbf{Internal validity.} The covered-versus-uncovered performance gap may be confounded by factors such as API documentation quality, task difficulty, or agent model capability. We mitigate this by using three independent agent models (spanning frontier, mid-tier, and local open-weights) and by grounding oracles in environment state rather than agent self-report.

\textbf{External validity.} Our 30-system sample (plus 7 for validation) may not represent all software categories, despite deliberate diversity across five categories and both commercial and open-source systems. Results may differ for mobile-only applications, embedded systems, or enterprise software with private APIs. The RQ3 validation is further limited to self-hosted open-source systems, which tend toward higher ACFC; the covered-versus-uncovered effect may differ in magnitude for low-ACFC commercial software.

\textbf{Reliability.} LLM-assisted capability extraction introduces potential inconsistency. We mitigate this by manually validating every extracted candidate against the guideline and by releasing the complete capability-to-interface mapping with per-score evidence, so that the scoring can be independently audited.

\section{Conclusion}
\label{sec:conclusion}

This paper articulated the GUI--API parity principle and introduced two instruments to operationalize it: Agent-Callable Feature Coverage (ACFC), which quantifies one measurable dimension of agent readiness, and the Agent Readiness Conformance (ARC), a multi-dimensional framework for assessing readiness along Accessibility, Discoverability, and Controllability.

Through an empirical study of 30 software systems (1,299 capabilities scored), we found a median ACFC of 0.55, with developer tools (median 0.69) leading and creative tools (0.41) trailing. Controllability is the primary readiness bottleneck: most systems achieve moderate API coverage ($\bar{A} = 2$) but lag on safety governance ($\bar{C} \leq 1$ in three of five categories; overall mean $C = 1.42$), revealing an ``ADC imbalance'' that binary coverage metrics cannot detect. Our validation shows that Accessibility is a necessary structural substrate for API-only agent execution (56 of 58 accessible tasks completed), while a controlled documentation ablation provides initial evidence that Discoverability affects success even among already-accessible capabilities. Inaccessible capabilities, used as negative controls, were not solved under API-only execution.
As computer-use agents increasingly bypass GUI restrictions, the question shifts from whether agents \emph{can} use software to whether software is \emph{designed} for agents. ACFC and ARC provide an initial measurement foundation for this transition; validating how Controllability metadata changes agent safety behavior remains important future work.


\balance
\bibliographystyle{IEEEtran}
\bibliography{references}

\end{document}